\documentclass[11pt,twoside]{article}
\usepackage{asp2010}
\usepackage{epsf}
\usepackage{lscape}
\usepackage{natbib}
\usepackage{graphicx}

\catcode`\@=11
\def\gsim{\ifmmode{\,\mathrel{\mathpalette\@versim>\,}}
    \else{$\,\mathrel{\mathpalette\@versim>}\,$}\fi}
\def\lsim{\ifmmode{\,\mathrel{\mathpalette\@versim<\,}}
    \else{$\,\mathrel{\mathpalette\@versim<}\,$}\fi}
\def\@versim#1#2{\lower 2.9truept \vbox{\baselineskip 0pt \lineskip
    0.5truept \ialign{$\m@th#1\hfil##\hfil$\crcr#2\crcr\sim\crcr}}}
\catcode`\@=12  
\def\av#1{\langle#1\rangle}

\newcommand\Mstar{M_*}

\newcommand\Msun{M_{\odot}}

\newcommand\rperi{r_{\rm peri}}
\newcommand\rvir{r_{\rm vir}}

\newcommand\avalpha{\langle\alpha\rangle}
\newcommand\dalpha{\delta\alpha}

\def\Mstarprime{M_*^{\,\prime}}

\def\Reprime{\Re^{\,\prime}}

\renewcommand\Re{R_{\rm e}}
\begin{document}
\bibliographystyle{asp2010}

\title{Dry mergers and the size evolution of early-type galaxies}

\author{Carlo~Nipoti
\affil{Dipartimento di Astronomia,
Universit\`a di Bologna, via Ranzani 1, I-40127 Bologna, Italy}}

\begin{abstract}
Dry (i.e. dissipationless) merging has been proposed as the main
driver of the observed size evolution of early-type galaxies (ETGs).
The actual role of this mechanism is questioned by the tightness of
the local stellar mass-size relation of ETGs. Combining this observed
scaling law with simple merging models, which should bracket
cosmologically motivated merging histories, we draw the following
conclusions:  1) local massive ETGs can have assembled at most $\sim
45\%$ of their stellar mass via dry mergers; 2) extreme fine tuning is
required for this to be the case.
\end{abstract}


{ \it    September 8, 2011}
\smallskip

In the hierarchical cosmological model galaxies are assembled by
mergers.  Early-type galaxies (ETGs) have old stellar populations
\citep[e.g.][]{Tho05}, which implies that, if they have undergone
mergers at $z \lsim 2$, these mergers must have been mostly ``dry''
(i.e., dissipationless, gas-poor, with little associated star
formation).  Present-day ETGs obey several empirical scaling laws,
whose existence represents a challenge for any theory of galaxy
formation.  Here we focus on the correlation between stellar mass
$\Mstar$ and effective radius $\Re$ ($\Re\propto\Mstar^{\,\beta}$,
with $\beta\sim 0.6-0.8$; e.g. \citealt{She03,Aug10}), and we address
the question of whether dry merging is able to reproduce this scaling
law.

An estimate of the effect of dry merging on the size of spheroids can
be obtained analytically on the basis of energy conservation and
virial theorem \citep{Hau78,Her93,Nip03}: under the assumption of
parabolic orbits, neglecting mass-loss and non-homology effects, one
gets $\Re\propto \Mstar^{\,\alpha}$, with $\alpha \sim 1-2$, depending
on the merger mass ratio $\xi$ \citep{Naa09}.  $N$-body
simulations---needed to take into account the effect of orbital
parameters, non-homology, mass loss, different distributions of stars
and dark matter (DM), and projection along the
line-of-sight---indicate that the analytic calculation gives a
reasonably good estimate of the {\it average} evolution of $\Re$ and
provide additional information on the {\it scatter} associated with
this average evolution (\citealt{Nip03,Boy06,Nip09b}, hereafter
N09b).The results of $N$-body simulations have been used to show that
{\it ETGs did not form via dry mergers of progenitors obeying the
  local scaling laws} \citep{Nip03}: in particular the dry-merger
remnants of such progenitors have too large $\Re$ as compared to real
ETGs.

\begin{table}[!ht]
\caption{Orbital parameters of the 13 minor-merger ($\xi=0.2$) simulations. }
\smallskip
\begin{center}
{\small
\begin{tabular}{ccccccccccc}
\tableline
\noalign{\smallskip}
\# & $\rperi/\rvir$ & $e$ & {~~~~~~} & \# & $\rperi/\rvir$ & $e$ & {~~~~~~} & \# & $\rperi/\rvir$ & $e$\\ 
\noalign{\smallskip}
\tableline
\noalign{\smallskip}
  1 &    0.000     &      1.00 & &  6 &    0.197     &  0.78 & & 11 &    0.203     &  0.89\\
  2 &    0.014     &      1.00 & &  7 &    0.200     &  0.94 & & 12 &    0.217     &  1.00\\
  3 &    0.054     &      1.00 & &  8 &    0.201     &  0.75 & & 13 &    0.339     &  1.00\\
  4 &    0.122     &      1.00 & &  9 &    0.201     &  0.81 & &    &              &\\
  5 &    0.195     &      1.06 & & 10 &    0.202     &  0.85 & &    &      &\\
\noalign{\smallskip}
\tableline
\end{tabular}
}
\end{center}
\end{table}

In the last years observations of $z\sim1.5-2.5$ ETGs have revealed
that many of these galaxies are much more compact than their local
counterparts, with $\Re$ smaller by a factor $3-5$ at similar stellar
mass \citep[e.g.][]{Dad05,Sar11}. Following this discovery, dry
merging (one of the few known mechanisms making galaxies less compact)
has become the most popular explanation for this size evolution of
ETGs \citep[e.g.][]{Kho06,Ose11}. An important question is whether dry
mergers can bring compact ETGs onto the $z=0$ scaling laws within
their scatter.  In an attempt to answer this question
\citet[][hereafter N09a]{Nip09a} and N09b combined $N$-body
simulations of galaxy mergers with accurate scaling laws of local ETGs
of the SLACS sample \citep{Bol08,Aug10}, showing that the tightness of
these scaling laws implies that extreme fine tuning is required in a
dry merging scenario and that ETGs can have accreted via dry mergers
at most $\sim$45\% of their stellar mass and $\sim$50\% of their total
(lensing) mass. It must be noted that the merging hierarchies
considered in N09ab are not based on a cosmological model, so the
natural step forward is to perform similar calculations, but for
cosmologically motivated merger histories (Nipoti et al., in
prep.). Here we present a first simple application of this approach.


In concordance $\Lambda$ cold dark matter cosmology the merging
history of DM halos is reasonably well constrained thanks to high
resolution cosmological $N$-body simulations such as Millenium I and
II \citep{Spr05,Boy09}. For these simulations \citet{Fak10} fit the
number of halo mergers as a function of merger mass ratio $\xi$,
redshift and halo mass. According to their fit, the mass-averaged mass
ratio is $\av\xi\sim 0.4$, independent of redshift and halo mass (see
Nipoti et al., in prep.).  Here we consider two idealized merger
histories that should bracket the behaviour of the cosmologically
motivated case with $\av\xi\sim 0.4$: a {\it major-merger hierarchy}
consisting of a sequence of $\xi=1$ mergers, and a {\it minor-merger
  hierarchy} consisting of a sequence of $\xi=0.2$ mergers. We
parametrize the effect of mergers by assuming that in each step of a
hierarchy $\Re\propto \Mstar^{\,\alpha}$, with $\alpha$ distributed
with average $\avalpha$ and standard deviation $\dalpha$ (hereafter
$\av{}$ and $\delta$ will always indicate average and standard
deviation, respectively).


\articlefigure[width =310 pt]{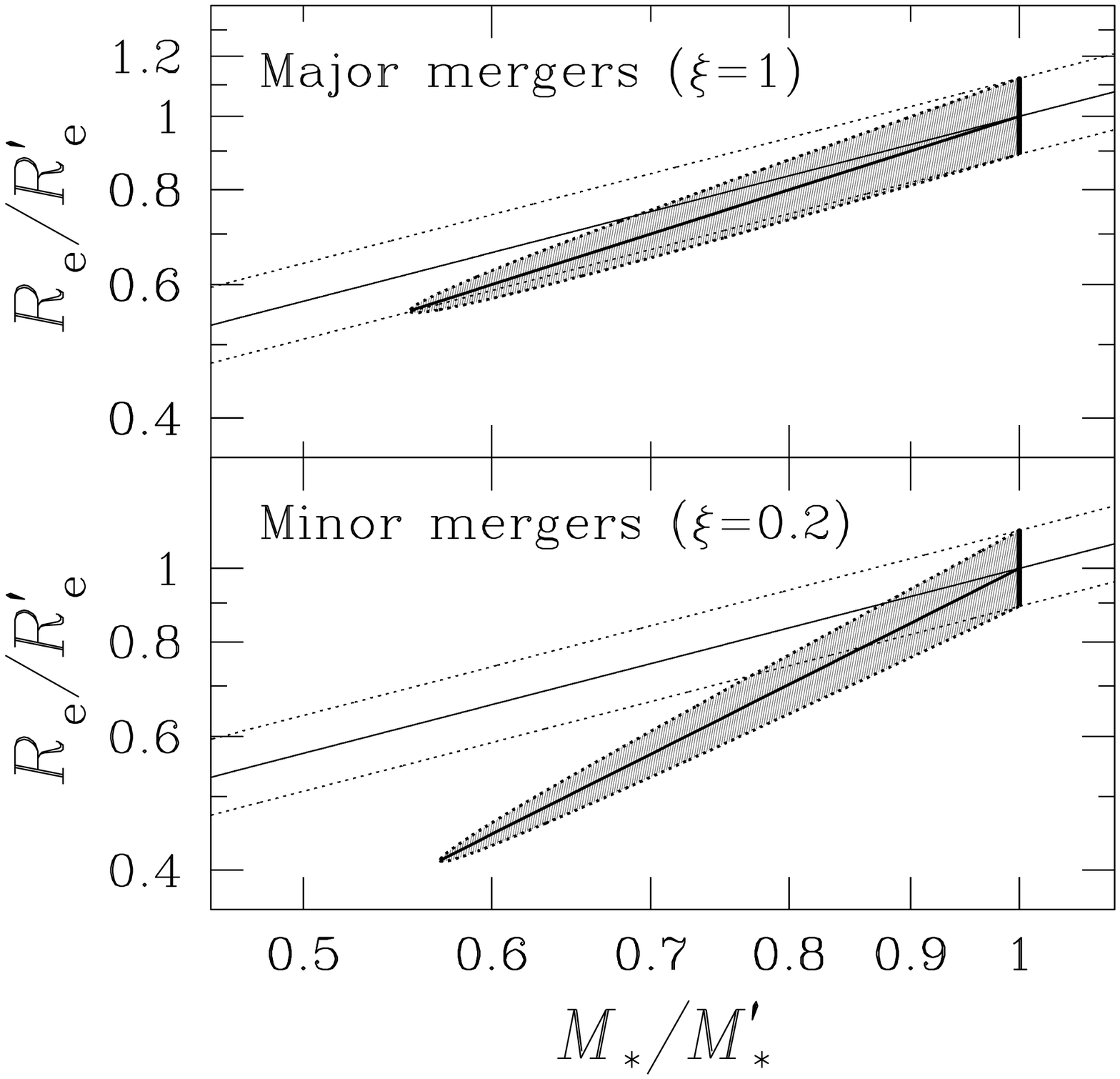}{fig:scatt}{The shaded
  areas represent the distributions in the $\Mstar$-$\Re$ plane of
  allowed major (upper panel) and minor (lower panel) dry-merging
  progenitors of local ETGs with stellar mass $\Mstarprime$ (vertical
  bars).  The upper and lower boundaries of the shaded areas
  correspond to 1-$\sigma$ of the distributions, while the solid lines
  within the shaded areas show their average values.  $\Reprime$ is
  the average value of $\Re$ for local ETGs with mass $\Mstarprime$.
  The diagonal solid and dashed lines indicate the local observed
  correlation with its 1-$\sigma$ scatter \citep{Aug10}.}

To estimate the distribution of $\alpha$ for $\xi=1$ we consider the
22 equal-mass mergers simulations of N09b, obtaining $\av\alpha \simeq
1.00$ and $\delta\alpha \simeq 0.18$.  For the $\xi=0.2$ case we get
$\av\alpha\simeq 1.60$ and $\delta\alpha\simeq 0.36$, based on a a new
set of 13 $N$-body dissipationless simulations (Nipoti et al., in
prep.). These simulations, run with the $N$-body code FVFPS
\citep{Lon03,Nip03}, model the encounter between two spherical
galaxies with the same distribution (model A in N09b), mass ratio 0.2,
size ratio 0.36, and different orbital parameters. For this set of
simulations we report in Table 1 the values of the eccentricity $e$
and of the pericentric radius $\rperi$ (in units of the main-halo
virial radius $\rvir$), which are distributed with $\av{e}\simeq
0.93$, $\delta{e}\simeq 0.10$, $\av{\rperi/\rvir}\simeq 0.17$ and
$\delta{(\rperi/\rvir)} \simeq 0.09$ (considering only bound orbits
the circularity $\eta$ is distributed with $\av{\eta}\simeq 0.53$ and
$\delta{\eta}\simeq 0.12$).  We note that the scatter in the orbital
parameters is comparable to that found in halo mergers in cosmological
$N$-body simulations by \citet[][]{Wet11}.

The obtained distributions of $\alpha$ can be used to estimate the
scatter in $\Re$ expected in a full merging hierarchy (see N09ab).  In
a single step of the hierarchy a galaxy increases its stellar mass
from $M_{*,i}$ to $M_{*,i+1}$ and its effective radius from $R_{{\rm
    e},i}$ to $R_{{\rm e},i+1}=(M_{*,i+1}/M_{*,i})^{\,\alpha}R_{{\rm
    e},i}$, with $\alpha$ distributed with $\avalpha$ and
$\delta\alpha$.  If we start with a population of galaxies all with
the same initial mass $M_{*,0}$, and $\log \Re$ distributed with
$\av{\log R_{{\rm e},0}}$ and $\delta \log R_{{\rm e},0}$, after $n$
merging steps with mass ratio $\xi$ we have a distribution of $\log
\Re$ with $\langle \log R_{{\rm e},n}\rangle=\av{\log R_{{\rm e},0}}+
\avalpha \log(M_{*,n}/M_{*,0})$ and $ \delta \log R_{{\rm e},n}=
\sqrt{(\delta \log R_{{\rm e},0})^2 +(\dalpha)^2\log(M_{*,n}/M_{*,0})
  \log (1+\xi)}, $ where $\avalpha$ and $\dalpha$ depend of $\xi$.
These expressions can be used to constrain the allowed properties of
the candidate dry-merging progenitors of descendant ETGs (see N09ab).
Here we consider the $\Re$-$\Mstar$ relation for the massive ETGs of
the SLACS sample as given in \citet{Aug10}: $\Re\propto\Mstar^{\,
  0.81}$ with intrinsic scatter $\delta\log\Re=0.05$ at fixed $\Mstar$
(diagonal solid and dashed lines in Fig.~\ref{fig:scatt}).  We find
that in our idealized, fixed-$\xi$ hierarchies, minor ($\xi=0.2$)
mergers introduce roughly as much scatter in $\Re$ as major ($\xi=1$)
mergers: in both merging hierarchies the distributions of allowed dry
merging progenitors (shaded areas in Fig.~\ref{fig:scatt}) are
relatively small and do not extend below $\Mstar/\Mstarprime \sim
0.55$ (where $\Mstar$ is the mass of the progenitor and $\Mstarprime$
is the mass of the descendant), which formally excludes a pure $\xi=1$
merging hierarchy.  Overall, the present analysis suggests that local
massive ($\Mstar\gsim 10^{11}\Msun$) {\it ETGs can have assembled at
  most $\sim 45\%$ of their mass growing with cosmologically motivated
  dry merger history} and that, for this to be the case, {\it extreme
  fine tuning is required}.  This result questions the scenario in
which compact $z \gsim 2$ ETGs evolve via dry mergers into local
massive ETGs. Though alternative explanations have been proposed,
including observational biases \citep[][]{Hop10,Man10}, QSO feedback
\citep[][]{Fan10} and decaying DM \citep{Fer09}, why high-$z$ galaxies
are so compact is still an open question.

\acknowledgments 

I am grateful to Matt Auger, Bernardo Nipoti and Tommaso Treu for
helpful discussions. I acknowledge support from the MIUR grant
PRIN2008 and the CINECA Award N. HP10C2TBYB (2011).





\bibliography{Nipoti_C} 

\end{document}